\documentstyle[aps,amsfonts,amssymb,eqsecnum,preprint]{revtex}
\tightenlines
\begin{document}

\sloppy

\draft

\title{Critical adsorption near edges}
\author{A. Hanke$^{1}$, M. Krech$^{2}$, F. Schlesener$^{1}$, and S. Dietrich$^{1}$}
\address{$^{1}$Fachbereich Physik, Bergische Universit\"at Wuppertal,\\
D-42097 Wuppertal, Federal Republic of Germany}
\address{$^{2}$Institut f\"ur Theoretische Physik, RWTH Aachen,\\
D-52056 Aachen, Federal Republic of Germany} 
\date{\today}
\maketitle
\begin{abstract}
Symmetry breaking surface fields give rise to nontrivial 
and long-ranged 
order parameter profiles for critical systems such as fluids,
alloys or magnets confined to wedges. We discuss the properties
of the corresponding universal scaling functions of the 
order parameter profile and the two-point 
correlation function and determine 
the critical exponents $\eta_{\parallel}$ and $\eta_{\perp}$ 
for the so-called normal transition.
\end{abstract}

\narrowtext

\section{Introduction}
\label{secI}

Advanced experimental techniques have emerged which allow one to endow
solid surfaces with stable geometrical structures which display a
well-defined design on the scale of nanometers or micrometers. Lateral
geometric structures can be formed by using various lithographic
techniques such as, e.g., holographic \cite{HKM92}, X-ray (LIGA)
\cite{Tol98}, soft \cite{XW98}, and nanosphere lithography \cite{Bur98}.
These microfabrication techniques provide routes to high-quality
patterns and structures with lateral dimensions down to tens of nm.
These structures are either periodic in one lateral direction
consisting of grooves with various shapes of the cross section
(e.g., wedgelike) or they display periodicity in both lateral
dimensions.

These man-made surfaces offer a wide range of possible applications if
they are exposed to fluids. In that case the surfaces act as a
template with designed topography which imposes specified lateral
structures on a fluid. For example in the context of microfluidics
\cite{Kni98,Gru99} these geometrical structures can be used as guiding
systems in order to deliver tiny amounts of valuable liquids to
designated analysis centers on a solid surface as part of microscopic
chemical factories \cite{Ser98}. Besides the numerous experimental
challenges associated with these systems there is also the theoretical
challenge to understand the corresponding highly inhomogeneous fluid
structures (and ultimately the flow dynamics) as well as to guide
the design on the basis of this insight \cite{Diet99}.

In view of this goal the study of the fluid structures in a single
wedgelike groove with opening angle $\alpha$ serves as a paradigmatic
first step. For $\alpha=\pi$ the geometry reduces to the well studied
case of a planar substrate. For decreasing values of $\alpha$ the
fluid is squeezed whereas for $\alpha>\pi$ a ridgelike solid
perturbation is projected into the fluid. Due to the unbound
accessible space for all values of $\alpha$ the bulk properties of the
fluid are unchanged by the presence of the wedge. (This is an
important difference as compared with pores in which the confinement
alters the bulk properties.) At low temperatures one finds pronounced
packing effects for the fluid particles near the corner of the wedge
which differ significantly from those encountered at planar surfaces
\cite{SD97,HSW98,NL90}. Even stronger effects occur if the wedge is exposed
to a vapor phase. In this case a liquid-like meniscus is formed at the
bottom of the wedge which undergoes a filling transition at a
temperature $T_{\alpha}$ {\em below\/} the wetting transition
temperature $T_W$ of the corresponding planar substrate
\cite{NKD92,Hau92,Lip98,RDN98}. It has been demonstrated that X-ray scattering
experiments at grazing incidence are capable of resolving such
interfacial structures \cite{Bus94,Li98}.

If the temperature is increased sufficiently above the triple point of
the fluid one encounters its bulk critical point $T_c$. This can be either
the liquid-vapor critical point of the fluid or the demixing critical
point if the fluid is a binary liquid mixture. The experience with
planar surfaces tells \cite{Bin83,Die86} that at bulk criticality the
confinement triggers interesting new surface critical phenomena. In
this case the local order parameter $\phi$ is perturbed near the
surface within a layer whose thickness is governed by the diverging
bulk correlation length $\xi_{\pm}(t\to 0) = \xi_0^{\pm} |t|^{-\nu}$
where $t=(T-T_c)/T_{c\,}$; $\nu$ denotes the universal bulk critical
exponent and $\xi_0^{\pm}$ is the nonuniversal amplitude above $(+)$
and below $(-)$ $T_c$. Thus at $T_c$ the perturbation due to the
surface intrudes deeply into the bulk.

In order to extend our knowledge of the structure of confined fluids
to these elevated temperatures we set out to investigate the
aforementioned surface critical phenomena in a wedge. Based on
analytic calculations \cite{Car83,GT84,BPP84,Lar86,AL95,KP96,DP97,KLT97} and
computer simulations 
\cite{PS98,PS99} the local critical behavior in a wedge has been
analyzed for the case that the corresponding planar surface exhibits
the so-called {\em ordinary\/} or {\em special\/} surface phase transitions
\cite{Bin83,Die86}. In the corresponding magnetic language the ordinary
phase transition corresponds to the case that the couplings between
the surface spins remain below the threshold value of the
multicritical special transition beyond which the surface can support
long-ranged order even above $T_c$ \cite{Bin83,Die86} and in
which there are no surface fields. The ordinary transition has also
been studied for different shapes of the confinement such as parabolic
ones \cite{PTI91,IPT93,KT95}. The main concern of these studies has been
critical edge exponents describing the leading critical behavior of
the order parameter near the corner where it differs from
that at planar surfaces and in the bulk. One finds that in general the
edge exponents depend on the opening angle $\alpha$.
These studies aim at describing magnetic systems in the absence of
surface fields as they are experimental accessible, e.g., by scanning
electron microscopy \cite{UCP91}.

However, the aforementioned surface and edge universality classes are
not applicable in the present context of confined fluids. For example,
in a binary liquid mixture one of the two species will prefer the
confining substrate more than the other which results in an effective
surface field acting on the order parameter given by the concentration
difference between the two species. Similar arguments hold for a
one-component fluid near liquid-vapor coexistence. These surface
fields give rise to another surface universality class, the so-called
{\em normal\/} transition \cite{Bin83,Die86}. It differs from the ordinary
and special transitions discussed above in that these symmetry
breaking surface fields generate a nontrivial order parameter profile
even if the bulk is in the disordered phase, i.e., for $T\ge T_c$.
For a planar surface this gives rise to the well studied so-called
{\it critical adsorption} phenomenon (see, e.g.,
Refs. \cite{Bin83,Die86,CD90,DS93,Die94,FD95,SL95}.) Here we extend the
corresponding field-theoretical analysis for the normal transition to the
wedge geometry under consideration. After presenting our model and
discussing general scaling properties (Sec.\,\ref{secII}) we analyze
the order parameter profile in Sec.\,\ref{secIII} and the structure
factor in Sec.\,\ref{secIV}.
By focusing on the critical temperature $T=T_c$ we are able to 
obtain analytical results. They are summarized in Sec.\,\ref{summary}.
Appendices \ref{appA} and \ref{appB} contain important technical
details needed in Secs.\,\ref{secIII} and \ref{secIV}.

Experimentally, wedges are characterized by a finite depth
$\Lambda$. Moreover, in many cases they are manufactured as a
laterally periodic array of parallel grooves and ridges. The
structural properties of a fluid exposed to such a periodic array
sensitively depends on the ratio $\Lambda / \xi$ of the depth and the
correlation length. For $\Lambda / \xi \ll 1$ 
the system will resemble a critical
fluid exposed to a mildly corrugated substrate which is expected to
give rise to corrections to the leading critical adsorption behavior at
a planar substrate. 
For $\Lambda / \xi \gg 1$ near the tip of a ridge and near the
bottom of a groove the presence of their periodic duplicates becomes
irrelevant so that in this limit the edge singularity of a single edge
with $\alpha>\pi$ and $\alpha<\pi$, respectively, will prevail. Our
present study deals with the limit $\Lambda \to \infty$ and
$\xi\to\infty$ taken such that $\Lambda / \xi \gg 1$. Certainly, in a
later stage it will be rather rewarding to study also the crossover
regime $\Lambda/\xi \approx 1$ 
in which the grooves and the ridges as well as
the way they are joined together play an equally important role. The
occurrence of the unbending transition in the case of wetting on
corrugated substrates \cite{SP98} indicates that the regime 
$\Lambda/\xi \approx 1$
might exhibit rather interesting new phenomena.


\section{Model and general results}
\label{secII}

Our investigation is based on the Ginzburg-Landau Hamiltonian ${\cal H}$
determining the statistical weight $\exp(-{\cal H})$ for a scalar order 
parameter $\phi$ which represents a critical system within
the Ising universality class. According to the wedge geometry,
${\cal H}$ is given as a sum of three contributions, i.e., 
${\cal H} = {\cal H}_b + {\cal H}_s + {\cal H}_e$, 
which refer to the bulk $(b)$, the surface $(s)$, and the edge
contribution $(e)$, respectively. In cylindrical coordinates 
${\bf x} = (r,\theta,{\bf x}_\|)$ (see Fig.\,\ref{fig_wedge})
the individual contributions are given by
\begin{mathletters}
\label{H}
\begin{equation} \label{Ha}
{\cal H}_b[\phi] = \int_0^\alpha d\theta 
\int_0^{\infty} dr \, r \int d^{d-2} x_\| \,
\Big\{ {1 \over 2}(\nabla \phi)^2 + {\tau \over 2} \phi^2 + {g \over 4!}
\phi^4 \Big\} \, \, ,
\end{equation}
\begin{eqnarray}
{\cal H}_s[\phi] =
\int_0^{\infty} dr \int d^{d-2} x_\| & & \Big\{ {c_1 \over 2}
[\phi(r,0,{\bf x}_\|)]^2 - h_{1\,} \phi(r,0,{\bf x}_\|) \label{Hb}\\
& & \, + \, {c_2 \over 2} [\phi(r,\alpha,{\bf x}_\|)]^2 - h_{2\,}
\phi(r,\alpha,{\bf x}_\|) \Big\} \, \, , \nonumber
\end{eqnarray}
\begin{equation} \label{Hc}
{\cal H}_e[\phi] = \int d^{d-2} x_\| \, \Big\{ {c_e \over 2}
[\phi(0,0,{\bf x}_\|)]^2 - 
h_{e\,} \phi(0,0,{\bf x}_\|) \Big\} \, \, ,
\end{equation}
\end{mathletters}
\noindent
where $d$ is the space dimension and 
${\bf x}_\|$ parameterizes the ($d-2$)-dimensional subspace parallel
to the edge along which the system is translationally invariant. 
The bulk Hamiltonian ${\cal H}_b$ given by
Eq.\,(\ref{Ha}) represents the standard Ginzburg-Landau 
$\phi^4$ model in the absence of external bulk fields.
The bulk parameter $\tau$ is proportional to the reduced temperature 
$t = (T - T_c) / T_{c\,}$, 
$g$ is the bulk coupling constant, 
and $\alpha$ denotes the opening angle.
The surface Hamiltonian ${\cal H}_s$ given by Eq.\,(\ref{Hb}) captures the
effect of the two semi-infinite planar surfaces forming the geometric
boundaries of the wedge located at $\theta = 0$ and $\theta = \alpha$,
respectively. Note that ${\cal H}_b$ and ${\cal H}_s$ take their
standard fixed-point form with a 
surface enhancement $c_i$ and a surface field $h_i$
for each of the surfaces $i = 1,2$. Cubic surface fields 
\cite{CD90} will be disregarded here. 
The surfaces meet at the opening angle $\alpha$
along the edge of the wedge which gives rise to the third contribution
${\cal H}_e$. The edge Hamiltonian ${\cal H}_e$ given by Eq.\,(\ref{Hc})
has the same structure as the surface contribution ${\cal H}_s$
and is characterized by an edge enhancement
$c_e$ and an edge field $h_e$ \cite{Die86}. The total Hamiltonian $\cal H$
constitutes a renormalizable model 
(see Subsec.\,IV\,A.3 in Ref.\,\cite{Die86}).

In this investigation we are exclusively concerned with the normal
transition which is characterized by nonzero values of
all surface and edge fields $h_1, h_2$, and $h_e$ and arbitrary values
of the surface enhancements $c_1$ and $c_2$ and of the edge enhancement
$c_e$. For nonzero fields $h_1$ or $h_2$ the 
system is {\em ordered} at any finite point even for $T > T_c$.
Asymptotically close to the critical point $T_c$ the universal 
properties of the corresponding order parameter profile 
are linked to the {\em critical adsorption fixed-point\/}
of the corresponding renormalization group description.
The ensuing scaling functions refer to the scaling limit 
$r \to \infty$ and $\xi \to \infty$, where the ratio 
$r / \xi$ is kept fixed forming a finite scaling variable.  
{\em At\/} the critical adsorption fixed-point the surface fields are
infinitely large so that the order parameter profile 
\begin{equation} \label{profile}
M(r,\theta,t;\alpha) \, \equiv \, 
\langle \phi(r,\theta,{\bf x}_\|) \rangle
\end{equation}
diverges at the surfaces $\theta = 0$ and $\theta = \alpha$
of the wedge according to a power law.
(We recall that such divergences refer to the
renormalization group fixed-point whereas actually the divergence of 
the order parameter profile is cut off at atomic distances
from the surfaces.)
Consequently, at this so-called normal transition the 
surfaces can differ at most with respect to the sign 
of the fields. 
In the following we assume that the surface fields have 
the same sign and we denote this configuration as $(+,+)$. 
Note that only ${\cal H}_b$ in conjunction with the 
aforementioned boundary conditions 
is left to determine the functional 
form of the order parameter profile.

Close to the critical point $T_c$ the order parameter profile 
takes the scaling form 
\begin{equation} \label{mpm}
M(r,\theta,t;\alpha) = a |t|^\beta 
P_\pm(r/\xi_\pm,\theta;\alpha) \, \, ,
\end{equation}
where $\xi_{\pm} = \xi_0^{\pm} |t|^{-\nu}$ with 
$t = (T - T_c) / T_c \gtrless 0$ is the correlation length
and $P_\pm(\zeta_\pm,\theta;\alpha)$ are the corresponding 
scaling functions with the scaling variable 
\begin{equation} \label{variable}
\zeta_\pm \, = \, r/\xi_\pm \, \, .
\end{equation}
The amplitude $a$ in Eq.\,(\ref{mpm})
is the nonuniversal amplitude of the bulk order
parameter $M_b = a |t|^\beta$, $T < T_c$, where $\beta$ is the 
corresponding bulk critical exponent. The scaling
functions $P_\pm$ are universal and have the limiting behavior 
$P_+(\zeta_+ \to \infty,\theta;\alpha) \to 0$ and
$P_-(\zeta_- \to \infty,\theta;\alpha) \to 1$, respectively. 
Note that $P_+$ and $P_-$ are universal but 
depend on the {\em definition} of the correlation length $\xi_\pm$,
because $\xi_\pm$ enters the scaling argument $\zeta_\pm$. In the opposite
limit $\zeta_\pm \to 0$, i.e., $T \to T_c$, the scaling functions $P_\pm$
exhibit short-distance singularities in form of power laws which
reflect the anomalous scaling dimension of the order parameter $\phi$:
\begin{equation} \label{Ppm}
P_\pm(\zeta_\pm \to 0,\theta;\alpha) = \tilde{\cal C}_\pm(\theta;\alpha)
\ \zeta_\pm^{-\beta/\nu} \, \, .
\end{equation}
The ratio $\beta/\nu$ of critical exponents has the values 
$1$ in $d = 4$, $\simeq 0.5168$ in $d = 3$ \cite{GZ85}, 
and $1/8$ in $d = 2$. 
Equation (\ref{Ppm}) implies a power law dependence on $r$ 
for the order parameter profile at criticality:
\begin{equation} \label{mpmTc}
M(r,\theta,t=0;\alpha) = a \, \tilde{\cal C}_\pm(\theta;\alpha)
(r/\xi_0^\pm)^{-\beta/\nu} \, \, .
\end{equation}
The scaling functions $\tilde{\cal C}_\pm(\theta;\alpha)$ appearing in
Eqs.\,(\ref{Ppm}) and (\ref{mpmTc}) are universal but depend on
the definition of the correlation length because they are derived
from $P_{\pm}$. (However, the product 
$\tilde{\cal C}_\pm (\xi_0^\pm)^{\beta/\nu}$
is invariant with respect to different choices for the 
definition of the bulk correlation length.)
In order to be specific we choose as the definition for $\xi$
the so-called {\em true} correlation
length which governs the decay of the two-point correlation function
in the bulk system \cite{FD95}. Equation (\ref{mpmTc}) implies the relation
\begin{equation} \label{CpCm}
\tilde{\cal C}_+(\theta;\alpha) / \tilde{\cal C}_-(\theta;\alpha) =
(\xi_0^+ / \xi_0^-)^{-\beta/\nu}
\end{equation}
between the scaling functions $\tilde{\cal C}_\pm$ involving the universal
ratio $\xi_0^+/\xi_0^-$. 

In the limit $\alpha = \pi$ the wedge geometry
coincides with the semi-infinite geometry. 
In this limit the 
universal scaling functions $\tilde{\cal C}_\pm$ 
reduce to
\begin{equation} \label{Cpmcpm}
\tilde{\cal C}_\pm(\theta;\alpha = \pi) = 
c_{\pm\,} (\sin \theta)^{-\beta/\nu}
\end{equation}
where the universal amplitudes $c_\pm$ with 
$c_+ / c_- = (\xi_0^+ / \xi_0^-)^{-\beta/\nu}$ 
govern the critical adsorption profile at a 
planar surface \cite{FD95} so that
\begin{equation} \label{mhalf}
M(r,\theta,t=0;\alpha=\pi) \, = \, 
M_{\infty/2}(z = r \sin \theta,t=0) \, = \, 
a \, c_{\pm\,} (z / \xi_0^\pm)^{-\beta/\nu}
\end{equation}
yields the order parameter profile $M_{\infty/2}$ in the semi-infinite system
at criticality. The profile $M_{\infty/2}(z_0,t=0)$ at a reference distance
$z_0$ from the surface can be used in order to construct the ratio
\begin{equation} \label{mpmmhalf}
{M(r,\theta,t=0;\alpha) \over M_{\infty/2}(z_0,t=0)} \, = \,
{\cal C}(\theta;\alpha) \, \rho^{-\beta/\nu} \, \, ,
\quad \rho = r / z_0 \, \, ,
\end{equation}
which is independent of the nonuniversal amplitudes $a$ and $\xi_0^\pm$.
Here we have introduced the length ratio 
$\rho = r/z_0$ and the scaling function
\begin{equation} \label{ratio}
{\cal C}(\theta;\alpha) \, \equiv \,  
\tilde{\cal C}_+(\theta;\alpha) / c_+ \, = \, 
\tilde{\cal C}_-(\theta;\alpha) / c_- \, \, .
\end{equation}
We note that ${\cal C}$ is not only universal but also 
independent of the definition for the 
correlation length because the lhs of Eq.\,(\ref{mpmmhalf})
and $\rho$ are expressed in terms of order parameter profiles 
and distances, respectively, without 
resorting to the notion of the correlation length.

In the limit that the normal distance 
$z = r \sin \theta$ from the surface of the wedge  
is much smaller than $r$, i.e., for 
$\theta \to 0$ with $r$ fixed, the order parameter profile 
reduces to the planar semi-infinite behavior 
\begin{equation} \label{planar}
M(r,\theta,t=0;\alpha) \, = \,
a \, c_{\pm\,} (z / \xi_0^\pm)^{-\beta/\nu}
\, \to \, a \, c_{\pm\,} \theta^{-\beta/\nu}
(r / \xi_0^\pm)^{-\beta/\nu} \, \, , \quad \theta \to 0 \, \, ,
\end{equation}
so that Eqs.\,(\ref{mhalf}) and (\ref{mpmmhalf}) imply
\begin{equation} \label{theta}
{\cal C}(\theta \to 0;\alpha) \, \to \, \theta^{-\beta/\nu} \, \, ,
\quad {\cal C}(\theta \to \alpha;\alpha) \, \to \, 
(\alpha-\theta)^{-\beta/\nu} \, \, ,
\end{equation}
where we have used the symmetry property
\begin{equation} \label{symmetry}
{\cal C}(\theta;\alpha) \, = \, 
{\cal C}(\alpha - \theta;\alpha) \, \, .
\end{equation}


\section{Order parameter profile at $T_{\lowercase{c}}$}
\label{secIII}

At the critical point the radial dependence of the order parameter profile
in the wedge is given by the power law $r^{-\beta/\nu}$ 
[see Eq.\,(\ref{mpmTc})]. The dependence 
of the profile on $\theta$ is captured by
the universal amplitude function ${\cal C}(\theta;\alpha)$ 
[see Eq.\,(\ref{mpmmhalf})], which in $d = 2$ is determined by
conformal invariance arguments (see Sec.\,\ref{secIIID}). 
In $d > 2$, however, one has to resort
to explicit field-theoretical calculations, 
assuming that the amplitude function ${\cal C}$,
as other universal quantities, is a smooth function of the spatial
dimension $d$. In the following we focus on the 
corresponding mean-field 
description, i.e., lowest order perturbation theory within the 
field-theoretical approach. The mean-field results become exact 
at the upper critical dimension $d_{\mbox{uc}} = 4$ of the field 
theory described by Eq.\,(\ref{H}), i.e., for $d \nearrow 4$.
In order to provide an estimate 
for the order parameter profile in $d = 3$ we shall use 
the exact results in $d = 2$ and the mean-field results 
in $d = 4$ for an 
interpolation scheme between $d = 2$ and $d = 4$, where the correct 
scaling arguments and short-distance singularities are 
implemented. For the order parameter profile studied here
we shall use the value $\beta / \nu \simeq 0.5168$
\cite{GZ85} in $d = 3$ instead of the mean-field value
$\beta / \nu = 1$. Thus only the amplitude function 
is treated in lowest order perturbation theory.


\subsection{Mean-field theory}
\label{secIIIA}

Within mean-field theory it is convenient to introduce 
the reduced profile 
\begin{equation} \label{mean}
m(r,\theta,\tau;\alpha) \, \equiv \, 
\sqrt{g/12} \ M(r,\theta,\tau;\alpha) \, \, ;
\end{equation}
$m(r,\theta,\tau;\alpha)$ is determined by the Euler-Lagrange equation
\begin{equation} \label{ELeq}
\left[{\partial^2 \over \partial r^2} + {1 \over r}{\partial \over \partial r}
+{1 \over r^2}{\partial^2 \over \partial \theta^2} \right] m = 
\tau \, m \, + \, 2 \, m^3
\end{equation}
for $r > 0$ and $0 < \theta <\alpha$.
The divergences of the profile $m$ near the surfaces and the edge
of the wedge account for the effect of the surface and edge parts
of the Hamiltonian in the $(+,+)$ configuration considered here. 
For $\tau = 0$ Eq.\,(\ref{ELeq}) implies for the universal 
amplitude function
\begin{equation} \label{C1}
{\cal C}(\theta;\alpha) \, = \, r \, m(r, \theta, \tau = 0; \alpha)
\end{equation}
[see Eq.\,(\ref{mpmmhalf}), here $\beta/\nu = 1$] the equation
\begin{equation} \label{C2}
{\partial^2 \over \partial \theta^2} \, {\cal C}(\theta;\alpha)
\, = \, - \, {\cal C}(\theta;\alpha) \, + \, 
2 \, [{\cal C}(\theta;\alpha)]^3 \, \, .
\end{equation}
Equation (\ref{C2}) leads to exact results for ${\cal C}$ 
in the limit $d \nearrow 4$.
Note that Eq.\,(\ref{C2}) can readily be solved 
using the order parameter profile obtained for a
{\em film geometry\/} with $(+,+)$ boundary conditions 
\cite{Kre97}. Specifically,
Eq.\,(\ref{C2}) is equivalent to the Euler-Lagrange equation for 
a film with thickness $L$ as  
given by Eq.\,(A3) in Ref.\,\cite{Kre97} if the rescaled temperature 
$\tau L^2$ there is replaced by $- \alpha^2$.  
The expression $\theta / \alpha$ for the wedge corresponds to 
$z / L$ for the film. The solution of Eq.\,(\ref{C2}) 
can be expressed by the Jacobian elliptic functions sn and dn,
\begin{mathletters} \label{pi}
\begin{equation} \label{m0pi}
{\cal C}(\theta;\alpha) \, = \, 
\frac{1}{\sqrt{1 - 2 k^2(\alpha)}} \, \, 
\frac{\mbox{dn}(\theta / \sqrt{1 - 2 k^2(\alpha)}; k(\alpha) )}
{\mbox{sn}(\theta / \sqrt{1 - 2 k^2(\alpha)}; k(\alpha) )} \, \, ,
\quad 0 \leq \alpha \leq \pi \, \, ,
\end{equation}
in terms of the modulus $k$ of the complete elliptic integral $K(k)$
of the first kind (see Appendix \ref{appA}) 
where $k = k(\alpha)$ is implicitly given by
\begin{equation} \label{k0pi}
\alpha = 2K(k) \sqrt{1 - 2 k^2} \, \, , 
\quad 1/2 \geq k^2 \geq 0 \, \, ,
\end{equation}
\end{mathletters}
\noindent
(see Eqs.\,(A12) and (A13) in Ref.\,\cite{Kre97}) and
\begin{mathletters} \label{2pi}
\begin{equation} \label{mpi2pi}
{\cal C}(\theta;\alpha) \, = \,  
\frac{1}{\sqrt{1 + k^2(\alpha)}} \, \,
\frac{1}
{\mbox{sn}(\theta / \sqrt{1 + k^2(\alpha)}; k(\alpha))} \, \, ,
\quad \pi \leq \alpha \leq 2 \pi \, \, ,
\end{equation}
where now $k = k(\alpha)$ is implicitly given by
\begin{equation} \label{kpi2pi}
\alpha = 2K(k) \sqrt{1 + k^2} \, \, ,
\quad 0 \leq k \leq k_{max} \simeq 0.90953 \, \, ,
\end{equation}
\end{mathletters}
\noindent
(see Eqs.\,(A14) and (A15) in Ref.\,\cite{Kre97}). In the
following we consider only the case $0 \leq \alpha \leq \pi$;
according to Eq.\,(\ref{2pi}) the extension to the case 
$\pi \leq \alpha \leq 2 \pi$ is straightforward.
Note that the parameterizations of the opening angle $\alpha$ 
as given by Eqs.\,(\ref{k0pi}) and (\ref{kpi2pi}) 
are monotonous functions of the modulus $k$ and therefore
the inverse function $k(\alpha)$ is uniquely defined but 
cannot be expressed in closed form in general. 
The special case $\alpha = \pi$ corresponds to $k(\alpha = \pi) = 0$ 
so that ${\cal C}(\theta;\alpha = \pi) = 1/\sin \theta$,
which reproduces
the mean-field result for the order parameter profile in the
semi-infinite geometry at the normal transition expressed 
in polar coordinates 
[see Eqs.\,(\ref{Cpmcpm}) and (\ref{ratio}), 
here $\beta/\nu = 1$].
The shape of $\cal C(\theta;\alpha)$ for $d = 2$ 
[see Eq.\,(\ref{function}) in Sec.\,\ref{secIIID}]
and $d = 4$ [see Eq.\,(\ref{pi})] 
for fixed opening angles $\alpha = \pi/2$ and $\alpha = \pi/4$
is shown in Fig.\,\ref{fig_C} where the divergences 
at $\theta = 0$ and $\theta = \alpha$ 
according to Eq.\,(\ref{theta}) have been split off. 
As an estimate for $d = 3$ we show in Fig.\,\ref{fig_C}
the linear interpolation between the corresponding curves for 
$d = 2$ and $d = 4$.


\subsection{Contour lines of the order parameter profile}
\label{secIIIB}

One of our motivations for the investigation of criticality in
the wedge geometry is the frequent
use of grooved substrates as templates to impose inhomogeneous structures
on liquids. In the wedge geometry the density (or concentration)
of the adsorbed critical fluid is inhomogeneous within 
planes perpendicular to the edge. The intersection of 
these planes with surfaces of constant order parameter 
$M(r,\theta,t=0;\alpha) = M_0$ renders the contour lines
$r = r(\theta;\alpha,M_0)$ of the order parameter profile,
which characterize the inhomogeneous shape of the 
order parameter configuration in the wedge.
Far away from the center
of the wedge the order parameter profile 
approaches the profile for a planar
substrate so that the contour lines asymptotically become straight
lines parallel to the surfaces forming the wedge 
(compare Fig.\,\ref{fig_wedge}). As expected,
the deviation of the contour lines
from these parallel straight lines is largest in the center of the 
wedge. The normalized order parameter profile defined by the lhs
of Eq.\,(\ref{mpmmhalf}) yields the contour lines 
\begin{equation} \label{contour}
\frac{r(\theta,\alpha)}{z_0} \, = \,
\left[ {\cal C}(\theta;\alpha) \right]^{\nu / \beta} \, \, ,
\end{equation}
where for a given contour line 
the reference distance $z_0$ is fixed by the asymptotic 
distance of this contour line from the wedge surface, i.e.,
by $M_0 = M_{\infty/2}(z_0,t=0)$ (see Fig.\,\ref{fig_wedge}).
Thus the angular dependence of {\em every\/} contour line exhibits 
the {\em same\/} universal shape in units of the 
corresponding asymptotic distance $z_0$.
Equations (\ref{theta}) and (\ref{contour})
imply that $r(\theta \to 0,\alpha) \sim \theta^{-1}$ 
independent of $d$ and $\alpha$. 

Within mean-field 
theory (for which $\beta/\nu = 1$) we obtain from Eq.\,(\ref{pi}) 
in the center of the wedge, i.e., for $\theta = \alpha/2$:
\begin{equation} \label{mcenter}
\frac{r(\theta = \alpha/2, \alpha)}{z_0} \, = \,
{\cal C}(\theta = \alpha/2;\alpha) \, = \,  
{\sqrt{1 - k^2(\alpha)} \over \sqrt{1 - 2 k^2(\alpha)}} \, \, ,
\quad d = 4 \, \, .
\end{equation}
According to Fig.\,\ref{fig_wedge} the distance $R$ 
of the intersection of the asymptotes from the edge is given by
$R = z_0/\sin(\alpha/2)$ which implies for the deviation 
$\Delta r = r - R$ of the contour line at the center of the wedge
\begin{equation} \label{Dr}
\Delta r = \left[{\sqrt{1 - k^2(\alpha)} \over \sqrt{1 - 2 k^2(\alpha)}}
- {1 \over \sin(\alpha/2)} \right] z_0 \, \, , \quad d = 4 \, \, .
\end{equation}
In order to discuss
Eq.\,(\ref{Dr}) we first consider a very wide wedge, i.e., 
the limit $\alpha \to \pi$ for which $k^2 \to 0$.
In this limit the deviation $\Delta r$
decays according to
[see Eqs.\,(\ref{k0pi}) and (\ref{Kkseries})]
\begin{equation} \label{Drpi}
{\Delta r \over z_0} \, = \, {2 \over 3} 
\Big( 1 - {\alpha \over \pi} \Big) +
\Big( {5 \over 6} - {\pi^2 \over 8} \Big) 
\Big(1 - {\alpha \over \pi} \Big)^2 \, \, + \, 
{\cal O}\Big( \Big(1 - {\alpha \over \pi} \Big)^3 \, \Big)
\, \, , \quad d = 4 \, \, .
\end{equation}
In the opposite limit of a very narrow wedge, i.e., for $\alpha \to 0$, 
the deviation $\Delta r$ of the contour line in the center 
{\em diverges\/} according to
\begin{equation} \label{Dr0}
\frac{\Delta r}{z_0} \, = \, 
\frac{\Omega}{\alpha} \, \, + \, {\cal O}(\alpha) 
\, \, , \quad d = 4 \, \, ,
\end{equation}
where $\Omega = \sqrt{2} K(1/\sqrt{2}) - 2 \simeq 0.622$. 
This shows that the
contour lines are expelled from the wedge into the bulk for small opening
angles. The overall behavior of $\Delta r$ for $0 \leq \alpha \leq \pi$
for $d = 2$ [see Eq.\,(\ref{Dr2d}) in Sec.\,\ref{secIIID}]
and $d = 4$ [see Eq.\,(\ref{Dr})] is shown in Fig.\,\ref{fig_dr}. 
Again, the curve for $d = 3$ represents the estimate
obtained by the linear interpolation between the curves for 
$d = 2$ and $d = 4$ (compare Fig.\,\ref{fig_C}).


\subsection{Distant wall corrections}
\label{secIIIC}

Far away from the center of the wedge the contour lines of the order 
parameter profile approach straight lines which are parallel to the
wedge surfaces (compare Fig.\,\ref{fig_wedge}).
This implies that in this limit the deviation of the order parameter 
profile from the corresponding
profile in a semi-infinite geometry becomes small. This small deviation can
be interpreted in terms of the so-called distant-wall correction which is
obtained from the {\em short-distance expansion} 
\cite{FY80,RJ82,Car86,Car90,ES94} of the order
parameter near, say, the surface $\theta = 0$
of the wedge. For any scaling
operator $\Psi$ the short-distance expansion at the 
normal transition can be written as
\begin{equation} \label{SDE}
\Psi(r,\theta,{\bf x}_\|) = \langle \Psi(r,\theta,{\bf x}_\|)
\rangle_{\infty/2\,} \left[1 + b_{\Psi} T_{\theta \theta}(r,0,{\bf x}_\|)
(r \theta)^d \, + \, {\cal O}(\theta^{d+2}) \, \right] \, \, ,
\end{equation}
where $T_{\theta \theta} = T_{\perp \perp}$ denotes the stress tensor
component {\em perpendicular} to the wall at $\theta = 0$ and $r \theta \ll
r \alpha$ is the small distance from the wall. 
For the order parameter profile
$M = \langle \phi \rangle$ we infer from Eq.\,(\ref{SDE})
\begin{equation} \label{SDEm}
M(r,\theta;\alpha) = M(r,\theta;\pi)\ [1 + B_{\phi}\ \theta^d + 
\dots \, ] \, \, , 
\end{equation}
in analogy to the short-distance expansion of the order parameter profile
in the film geometry \cite{FdG78}, where 
$B_{\phi} = b_{\phi} \langle T_{\theta \theta} \rangle r^d$. 
From the general expression of the stress tensor at criticality
(see Eq.\,(3.1) in Ref.\,\cite{ES94}) we obtain in polar coordinates
\begin{equation} \label{Tthth}
T_{\theta \theta} =
{1 \over 2r^2} \left({\partial \phi \over \partial \theta} \right)^2
- {1 \over 2} \left({\partial \phi \over \partial r}\right)^2
- {1 \over 2} \left( \nabla_\| \phi \right)^2 - {g \over 24} \phi^4
+ {d - 2 \over 4 (d - 1)}\ \left[{\partial^2 \over \partial r^2} \phi^2
+ \Delta_\| \phi^2 \right]
\end{equation}
up to contributions of the order $g^3$, where $\nabla_\|$ and $\Delta_\|$
denote the components of the gradient and the Laplacian along the edge,
respectively. 

Within mean-field theory the thermal average of 
Eq.\,(\ref{Tthth}) is obtained by setting $d = 4$ and replacing 
$\phi$ by $\langle \phi \rangle = \sqrt{12/g}\ m$ 
[see Eqs.\,(\ref{mean}) and (\ref{C1})]. 
Since the result is independent of $\theta$ we can use 
the fact that the order parameter profile is
symmetric around the midplane $\theta = \alpha/2$
in order to obtain
\begin{equation} \label{Tththav}
\langle T_{\theta \theta} \rangle = {6 \over g r^4}
\Big[ [{\cal C}(\alpha/2;\alpha)]^2 - 
[{\cal C}(\alpha/2;\alpha)]^4 \Big] \, \, , \quad d = 4 \, \, .
\end{equation}
From Eq.\,(\ref{pi}) we thus find
\begin{equation} \label{Tththk}
\langle T_{\theta \theta} \rangle \, = \, - \, {6 \over g r^4} \, 
{k^2(\alpha) [1 - k^2(\alpha)] \over [1 - 2k^2(\alpha)]^2} 
\, \, , \quad d = 4 \, \, ,
\end{equation}
which exactly corresponds to $\langle T_{zz} \rangle$ in the film geometry
with $(+,+)$ boundary conditions (see the first line of Eq.\,(3.4) 
in Ref.\,\cite{Kre97}
for $y = -1$). If Eq.\,(\ref{Tththk}) is inserted into Eq.\,(\ref{SDE}) for
$\Psi = \phi$ the structure of the distant-wall correction as given by
Eq.\,(\ref{SDEm}) is confirmed.
However, for a complete identification the
amplitude $b_{\phi}$ must be determined. 
Within the present mean-field approach
Eq.\,(\ref{SDEm}) can directly be verified by expanding Eq.\,(\ref{pi}) for
$\theta \to 0$ with $\alpha$ fixed. 
From the expansions of the Jacobian elliptic functions
[see Eq.\,(\ref{sndnseries})] we obtain
\begin{equation} \label{SDEmMF}
m(r,\theta;\alpha) = m(r,\theta;\pi)\
\left[1 + {k^2(\alpha) [1 - k^2(\alpha)] \over 
[1 - 2k^2(\alpha)]^2} \left({\theta^4 \over 10}
- {\theta^6 \over 105} \right) \, + \, {\cal O}(\theta^8) \right]
\, \, , \quad d = 4 \, \, ,
\end{equation}
from which one can read off $b_{\phi} = -g/60$ within mean-field theory.
The next-to-leading correction ${\cal O}(\theta^6)$ in Eq.\,(\ref{SDEmMF})
is generated by the operator
$\Delta_r T_{\theta \theta}$ where $\Delta_r$ denotes the radial part
of the Laplacian in polar coordinates $(r,\theta,{\bf x}_\|)$.
In general the corresponding correction is 
${\cal O}(\theta^{d+2})$ as indicated in Eq.\,(\ref{SDE}). In
principle $\Delta_\| T_{\theta \theta}$ also yields a contribution to the
short-distance expansion, but one has
$\Delta_\| \langle T_{\theta \theta} \rangle = 0$ 
due to the translational invariance along the ($d - 2$)-dimensional edge. 
We note that in the {\em film geometry\/} the counterpart of 
the operator $\Delta_r T_{\theta \theta}$ in the wedge does 
{\em not\/} contribute to the next-to-leading distant-wall correction
of the order parameter profile due to the translational 
invariance in  the film. 
Therefore in the film the next-to-leading correction
is given by the operator $(T_{\perp \perp})^2$, 
which gives rise to ${\cal O}(\theta^{2 d})$ corrections only.
The latter operator appears also in the wedge geometry and gives 
rise to the ${\cal O}(\theta^8)$ corrections 
indicated in Eq.\,(\ref{SDEmMF}).

The distant-wall correction has an interesting geometric 
interpretation. The contour lines of the order parameter profile in
the wedge and in the semi-infinite geometry, respectively,
belonging to the same value $M_0$ of the order parameter
are displaced by an amount $\delta(y;\alpha)$
where $\delta + z_0 = z = r \sin \theta$ and the distance $y$ is
related to $\theta$ by (compare Fig.\,\ref{fig_wedge})
\begin{equation} \label{y}
y \, = \, y(\theta,\alpha) \, = \, 
r(\theta,\alpha) \cos \theta \, - \, 
[R(\alpha) + \Delta r(\alpha)] \cos(\alpha/2) \, \, .
\end{equation}
Within mean-field theory one obtains for the 
above choice of $M_0$ 
[see Eq.\,(\ref{contour}), here $\beta/\nu=1$]
and using $r / z_0 = {\cal C}(\theta;\alpha)$
\begin{equation} \label{delta}
\frac{\delta (y(\theta,\alpha);\alpha)}{z_0} \, = \, 
{\cal C}(\theta;\alpha) \sin(\theta) \, - \, 1 \, = \, 
{m(r,\theta;\alpha) \over m(r,\theta;\pi)} \, - \, 1
\, \, , \quad d = 4 \, \, .
\end{equation}
According to Eqs.\,(\ref{SDEm}) and (\ref{delta}) the ratio
$\delta(y(\theta,\alpha);\alpha)/z_0$ as function of $\theta$
represents the distant-wall correction of the order parameter
profile. Its leading terms can be read off from Eq.\,(\ref{SDEmMF}). 
For general $d$ one has [see Eq.\,(\ref{contour})]
\begin{equation} \label{eins}
\frac{\delta(y(\theta,\alpha);\alpha)}{z_0} \, = \, 
[{\cal C}(\theta,\alpha)]^{\nu/\beta} \, \sin(\theta) \, - \, 1
\end{equation}
so that 
$\delta / z_0 \to (1/6) [(\pi/\alpha)^2 - 1] \, \theta^2$
for $d = 2$ and $\theta \to 0$
[see Eq.\,(\ref{function}) in Sec.\,\ref{secIIID}].
The behavior of 
$\delta(y,\alpha)$ as function of $y$ for fixed opening
angles $\alpha = \pi / 2$ and $\alpha = \pi / 4$
is shown in Fig.\,\ref{fig_delta}
for $d = 2$ [see Eq.\,(\ref{function})]
and $d = 4$ [see Eq.\,(\ref{delta})]. 
In order to obtain an estimate for $d = 3$ we introduce
\begin{equation} \label{interpol}
f(\theta;\alpha) \, = \, 
\frac{\delta(y(\theta,\alpha);\alpha)}{z_0} \
\left( \frac{2 \theta}{\alpha} \right)^{-d} \, \, .
\end{equation}
One has $\delta / z_0 \sim \theta^d$ for $\theta \to 0$ or
$\sim y^{-d}$ for $y \to \infty$ due to
$y(\theta \to 0, \alpha) = r(\theta \to 0, \alpha) \sim \theta^{-1}$
[see Eq.\,(\ref{y})]. Therefore 
$f(\theta;\alpha)$ tends for $\theta \to 0$
to a constant value (depending on $d$).
For $y = 0$ one has $2 \theta / \alpha = 1$ and
$f = \Delta r \sin(\alpha/2) / z_0$
(compare Figs.\,\ref{fig_wedge} and \ref{fig_dr}).
A reasonable estimate for $\delta(y;\alpha)$ 
in $d = 3$ can be obtained by interpolating 
$f(\theta;\alpha)$ linearly between 
$d = 2$ and $d = 4$ and using this approximation for $f$ in 
\begin{equation} \label{d3}
\frac{\delta(y(\theta,\alpha);\alpha)}{z_0} \, = \, f(\theta;\alpha)
\left( \frac{2 \theta}{\alpha} \right)^{3} \, \, ,
\quad d = 3 \, \, .
\end{equation}
The curves for $d = 3$ in Fig.\,\ref{fig_delta} correspond to the 
rhs of Eq.\,(\ref{d3}) for fixed values $\alpha = \pi / 2$
and $\alpha = \pi / 4$.
For $y = 0$ one has $2 \theta / \alpha = 1$ so that  
for $y = 0$ the linear interpolation of $f$ implies 
a linear interpolation of $\delta$
[see Eq.\,(\ref{interpol})].


\subsection{Exact results in $d = 2$}
\label{secIIID}

At criticality systems exhibit not only scale invariance but, 
more generally, {\em conformal} invariance \cite{Car86}.
This property is particularly useful in $d = 2$,
where the {\em large conformal group} provides mappings between many different
geometries \cite{Car86}. In higher spatial 
dimensions only M\"obius transformations are
available as conformal mappings (small conformal group) which map geometries
bounded by planes and spheres onto other geometries bounded by planes and
spheres. In $d = 2$ a wedge with
opening angle $\alpha$ can be conformally mapped
onto a half-plane, for which the profile is known. This leads to \cite{BE85}
\begin{equation} \label{phiwedge}
M(r,\theta, t = 0; \alpha) \equiv
\langle \phi \rangle_{wedge} = A\ (\pi/\alpha)^{\beta/\nu} \
[r \sin(\theta \pi / \alpha)]^{-\beta/\nu}
\end{equation}
in the wedge, where $\beta / \nu = 1/8$ within the Ising universality class
considered here. Note that conformal invariance {\em completely} determines
the universal amplitude function ${\cal C}(\theta;\alpha)$ 
[see Eq.\,(\ref{mpmmhalf})] as
\begin{equation} \label{function}
{\cal C}(\theta;\alpha) \, = \, (\pi/\alpha)^{\beta/\nu} \
[\sin(\theta \pi / \alpha)]^{-\beta/\nu} \, \, , \quad d = 2 \, \, ,
\end{equation}
which would remain unspecified by scale 
invariance considerations alone.

Using Eqs.\,(\ref{function}) and (\ref{contour}) we obtain the relation 
\begin{equation} \label{d2}
\frac{r(\theta;\alpha)}{z_0} \, = \, 
\frac{\pi/\alpha}{\sin(\theta\pi/\alpha)} \, \, , \quad d = 2 \, \, ,
\end{equation}
for the contour lines of the order parameter profile. The
deviation $\Delta r$ from the asymptotic contour lines at the center of
the wedge, i.e., $\theta = \alpha/2$, is defined as shown in 
Fig.\,\ref{fig_wedge}. We obtain
\begin{equation} \label{Dr2d}
\frac{\Delta r}{z_0} \, = \, {\pi \over \alpha} \, - \, 
{1 \over \sin(\alpha/2)}  \, \, , \quad d = 2 \, \, .
\end{equation}
In the limit $\alpha \to \pi$ Eq.\,(\ref{Dr2d}) yields the expansion
\begin{equation} \label{Dr2dpi}
{\Delta r \over z_0} \, = \, \Big(1 - {\alpha \over \pi} \Big) +
\Big(1 - {\pi^2 \over 8} \Big) \Big(1 - {\alpha \over \pi} \Big)^2 
\, + \, \, {\cal O} \Big( \Big(1 - {\alpha \over \pi} \Big)^3 \, \Big) \, \, ,
\quad d = 2 \, \, ,
\end{equation}
which is the analogue of Eq.\,(\ref{Drpi}) for $d = 2$.
For $\alpha \to 0$ we obtain from Eq.\,(\ref{Dr2d})
\begin{equation} \label{Dr2d0}
\frac{\Delta r}{z_0} \, = \, \frac{\pi - 2}{\alpha} 
\, \, + \, {\cal O}(\alpha) \, \, , \quad d = 2 \, \, ,
\end{equation}
which is the analogue of Eq.\,(\ref{Dr0}). Note that Eqs.\,(\ref{Drpi}) and
(\ref{Dr2dpi}) and Eqs.\,(\ref{Dr0}) and (\ref{Dr2d0}), respectively,
are very similar even in a quantitative sense, i.e.,
the dependence on the spatial dimension $d$ is weak. 

We close this subsection with a discussion of the short-distance expansion
of the order parameter according to Eq.\,(\ref{SDE}) 
for the case $d = 2$. The thermal
average of the stress tensor $\langle T \rangle$ in the wedge can be
obtained by means of a conformal mapping from 
the half-plane \cite{KP96} and reads in polar coordinates 
\begin{mathletters} 
\label{Tpolneu}
\begin{equation} \label{Tpola}
\langle T_{rr} \rangle = -\langle T_{\theta \theta} \rangle =
{c \over 24 \pi} \left[ \left({\pi \over \alpha}\right)^2 - 1 \right]
{1 \over r^2} \, \, ,
\end{equation}
\begin{equation} \label{Tpolb}
\langle T_{r \theta} \rangle = 
\langle T_{\theta r} \rangle = 0 \, \, ,
\end{equation}
\end{mathletters}
\noindent
where $c$ is the conformal anomaly number.
The distant-wall correction for the order parameter profile according to
Eq.\,(\ref{SDEm}) can be directly calculated from the expansion of
Eq.\,(\ref{phiwedge}) for small $\theta$ with the result
\begin{equation} \label{SDEm2d}
M(r,\theta, t = 0;\alpha) = 
M(r,\theta, t = 0;\pi)\ \Big\{ 1 + {\beta \over 6\nu} \Big[
\Big( {\pi \over \alpha} \Big)^2 - 1 \Big] \theta^2 + 
{\cal O}(\theta^4) \Big\} \, \, .
\end{equation}
From Eqs.\,(\ref{SDE}), (\ref{Tpolneu}), and (\ref{SDEm2d}) we
obtain $b_{\phi} = -(4\pi / c) \, \beta/\nu$ in $d = 2$.
For the Ising universality class considered here ($c = 1/2$ and 
$\beta/\nu = 1/8$) we have $b_{\phi} = -\pi$. 
Finally, we note that the next-to-leading
distant-wall correction ${\cal O}(\theta^4)$ 
indicated in Eq.\,(\ref{SDEm2d})
stems from $\Delta_r T_{\theta \theta}$ and 
$(T_{\theta \theta})^2$
because both operators have the same scaling dimension four.


\section{Static structure factor}
\label{secIV}

The static structure factor $S(|{\bf q}|)$ in a bulk system is
given by the Fourier transform of the two-point correlation function
$G(|{\bf r} - {\bf r}'|)$, where ${\bf q}$ denotes the momentum transfer and
${\bf r}$ and ${\bf r}'$ are positions in space. Due to translational
invariance and isotropy of the bulk the moduli of the momentum transfer and
the distance in space remain as the only arguments in Fourier space and real
space, respectively. In the wedge geometry translational invariance and
isotropy only hold in the ($d - 2$)-dimensional subspace along the edge so
that the position dependence of $G$ is more complicated.
Apart from modifications of the eigenmode spectrum caused
by different boundary conditions we follow the derivation of $G$
for a wedge as given in Ref.\,\cite{Car83}.


\subsection{Mean-field theory}
\label{secIVA}

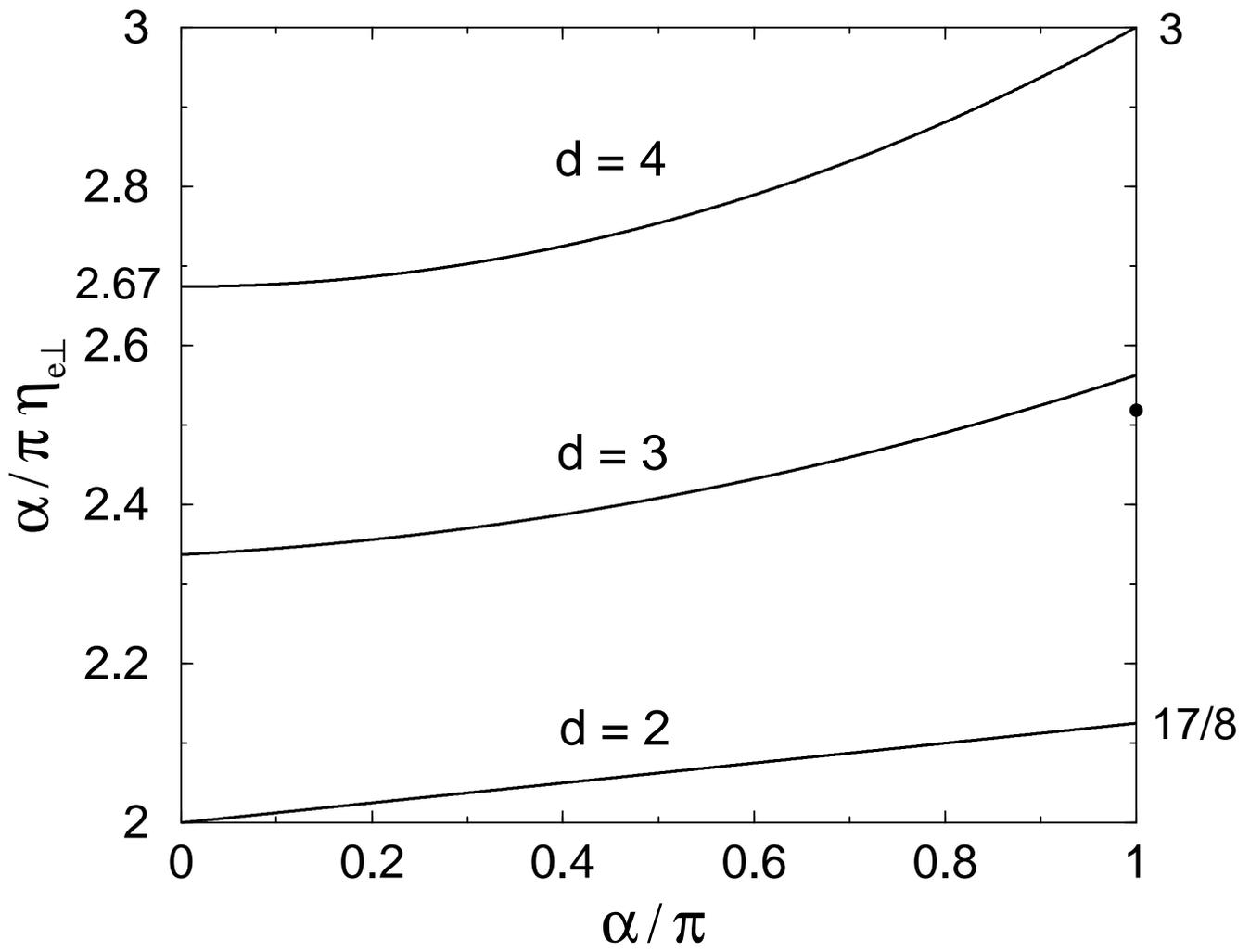
\begin{figure}
\caption{Universal dependence of the edge
exponent $\eta_{e \perp}(\alpha)$ [see Eq.\,(\ref{perp})]
on the opening angle $\alpha$, with the overall 
$1/\alpha$ dependence split off. 
The curve for $d = 3$ shows the linear interpolation
between the exactly known curves for $d = 2$ 
[see Eq.\,(\ref{eta2d})] and $d = 4$ [see Eq.\,(\ref{etaperp})].
The dot for $\alpha / \pi = 1$ shows the known value 
$\eta_{\perp} = (\eta + \eta_{\parallel})/2 \simeq 2.518$
corresponding to a planar surface in $d = 3$.
The small difference between the dot and the value 
of the interpolated curve for $d = 3$ is a measure of the 
uncertainty associated with the aforementioned linear 
interpolation scheme.}
\label{fig_eta}
\end{figure}

Starting from the formal definition 
$G({\bf r};{\bf r}') \equiv 
[\delta m({\bf r}) / \delta h({\bf r}')]_{h=0\,}$, 
where $m({\bf r})$ denotes the order
parameter profile and $h({\bf r})$ is an external 
spatially varying field, we obtain
\begin{equation} \label{Geq}
- \Delta G({\bf r};{\bf r}') + 6 m^2({\bf r}) G({\bf r};{\bf r}')
= \delta ({\bf r} - {\bf r}')
\end{equation}
at the critical point and within mean-field theory. 
For ${\bf r} = (r, \theta, {\bf x}_\|)$ 
we apply a Fourier transform with respect to ${\bf x}_\|$ and
define $S({\bf p};r,\theta;r',\theta')$ by \cite{Car83}
\begin{equation} \label{Sdef}
G(r,\theta,{\bf x}_\|;r',\theta',{\bf x}_\|') = \int {d^{d-2} p \over
(2\pi)^{d-2}} \, S({\bf p};r,\theta;r',\theta') 
\exp[i{\bf p} ({\bf x}_\| - {\bf x}_\|')] \, \, .
\end{equation}
From Eqs.\,(\ref{Geq}) and (\ref{Sdef}) and by using 
$m({\bf r}) = {\cal C}(\theta;\alpha) r^{-1}$ 
[see Eq.\,(\ref{C1})] one obtains
\begin{equation} \label{Seq}
-\left[{\partial^2 \over \partial r^2} + {1 \over r} {\partial \over
\partial r} + {1 \over r^2} {\partial^2 \over \partial \theta^2} - p^2
\right] S + {6 \over r^2} \ [{\cal C}(\theta;\alpha)]^2 \ S 
= {1 \over r} \delta(r - r') \delta(\theta - \theta') \, \, .
\end{equation}
We solve Eq.\,(\ref{Seq}) in terms of eigenfunctions to a spectrum $E$ of
eigenvalues for the operator on the left hand side of Eq.\,(\ref{Seq}). Each
eigenfunction can be written as a product of a radial part $R(r)$ and an
angular part $\psi(\theta)$. For the shifted spectrum $\kappa^2 = E -
{\bf p}^2$ the corresponding eigenvalue equations are given by
\begin{equation} \label{Rr}
R_n^{\,''}(r) + r^{-1} R_n^{\,'}(r) + 
(\kappa^2 - \lambda_{n\,} r^{-2}) R_n(r) = 0
\end{equation}
and
\begin{equation} \label{psit}
-\psi^{''}_n(\theta) + 6\ [{\cal C}(\theta;\alpha)]^2 \ \psi_n(\theta) 
= \lambda_{n\,} \psi_n(\theta) \, \, ,
\end{equation}
where Eq.\,(\ref{Rr}) is a Bessel equation with parameter $\kappa$ and
Eq.\,(\ref{psit}) is a Lam{\'e} equation. In general the discrete
eigenvalues $\lambda_n$ cannot be given in closed form
(see Appendix \ref{appB} for details). 
Apart from the specific spectrum $\{ \lambda_n \}$ and the corresponding 
eigenfunctions $\{ \psi_n \}$ one can represent the solution of 
Eq.\,(\ref{Seq}) in the same form as in Ref.\,\cite{Car83}, i.e.,
\begin{eqnarray} \label{Sspectral}
S({\bf p};r,\theta;r',\theta') &=& {2 \over \alpha} \sum_{n=3}^\infty
\int_0^\infty d\kappa \ \kappa \ {J_{\sqrt{\lambda_n}}(\kappa r)
J_{\sqrt{\lambda_n}}(\kappa r') \over \kappa^2 + {\bf p}^2}\
\psi_n(\theta) \psi_n(\theta') \nonumber\\[2mm]
&=& {2 \over \alpha} \sum_{n=3}^\infty I_{\sqrt{\lambda_n}}(p r_<)
K_{\sqrt{\lambda_n}}(p r_>)\ \psi_n(\theta) \psi_n(\theta')
\end{eqnarray}
where $p = |{\bf p}|$, $r_< = \min (r,r')$, and $r_> = \max (r,r')$. 
In order to determine the decay of the two-point correlation function
in real space 
we insert Eq.\,(\ref{Sspectral}) into Eq.\,(\ref{Sdef})
and proceed along the lines described in Ref.\,\cite{Car83}. The decay of $G$ 
away from the edge is governed by the critical exponent
$\eta_{e \perp}(\alpha)$ according to 
\begin{equation} \label{perp}
G(r,\theta,{\bf x}_\|; r'\to \infty, \theta', {\bf x}'_\|) 
\, \sim \, \frac{1}{(r')^{d-2+\eta_{e \perp}(\alpha)}}
\end{equation}
and as in Ref.\,\cite{Car83} we find from Eq.\,(\ref{Sdef})
that in the limit $r'/r \to \infty$
the smallest eigenvalue in the spectrum $\{ \lambda_n \}$ determines 
the exponent $\eta_{e \perp}(\alpha)$.
With the notation used in Appendix B
the smallest eigenvalue is $\lambda_3$ and in accordance with
Ref.\,\cite{Car83} we find $\eta_{e \perp}(\alpha) = \sqrt{\lambda_3}$. 
Along the edge, i.e., in
the limit $|{\bf x}_\| - {\bf x}'_\|| \to \infty$ for $r,r' > 0$ the decay
of $G$ is governed by the critical exponent
$\eta_{e \|}(\alpha)$ according to 
\begin{equation} \label{para}
G(r,\theta,{\bf x}_\|; r', \theta', {\bf x}'_\| \to \infty) \, \sim \, 
\frac{1}{|{\bf x}_\| - {\bf x}'_\||^{d-2+\eta_{e \parallel}(\alpha)}}
\, \, .
\end{equation}
In agreement with the general scaling relation 
\begin{equation} \label{eta}
\eta_{e \perp}(\alpha) \, = \, \frac{\eta + \eta_{e \|}(\alpha)}{2}
\, \, , \quad 2 \leq d \leq 4 \, \, ,
\end{equation}
we find $\eta_{e \|}(\alpha) = 2\eta_{e \perp}(\alpha)$ within 
mean-field theory (for which $\eta = 0$). 
From the approximate evaluation of the spectrum
$\{ \lambda_n \}$ [see Eq.\,(\ref{epsn}) in Appendix B]
we obtain
\begin{equation} \label{etaperp}
\eta_{e \perp}(\alpha) \, = \, \sqrt{\lambda_3} \, \approx \,  
{3\pi \over \alpha}\sqrt{1 - {4 \alpha \over 3\pi^{2}} \,
\zeta(\alpha/2) + {2\alpha^2 \over 9 \pi^4} 
\left(\pi^2 + {g_2(\alpha) \over 24} \alpha^2 - 
2\zeta^2(\alpha/2) \right)}
\end{equation}
where $\zeta(u)$ is the Weierstrass $\zeta$-function 
[see Eq.\,(\ref{Wzeta})] and 
\begin{equation} \label{g2}
g_2(\alpha) \, = \,
\frac{4 k^2(\alpha) [1-k^2(\alpha)]}{[1-2k^2(\alpha)]^2} 
\, + \, \frac{4}{3}
\end{equation}
[see Eq.\,(\ref{WPwedge})].
In the limit $\alpha \to \pi$ corresponding to a planar surface
Eq.\,(\ref{etaperp}) becomes exact
and one obtains $\eta_\perp \equiv \eta_{e \perp}(\pi) = 3$ in
accordance with the mean-field result for the semi-infinite geometry at the
normal transition \cite{Bin83}. The dependence of $\eta_{e \perp}$ 
on $\alpha$ is displayed in Fig.\,\ref{fig_eta} for $d = 2$
[see Eq.\,(\ref{eta2d}) in Sec.\,\ref{secIVB}] and 
$d = 4$ [see Eq.\,(\ref{etaperp})], where the overall 
$1/\alpha$ dependence has been split off.
The estimate for $d = 3$ is obtained by interpolating linearly between
the curves
for $d = 2$ and $d = 4$ (compare Figs.\,\ref{fig_C} and \ref{fig_dr}).
The exponent $\eta_{e \perp}(\alpha)$ increases 
monotonically as $\alpha$ decreases. This 
implies that at the normal transition
the correlation function decays more rapidly 
away from an edge ($\alpha < \pi$)
than away from a planar surface ($\alpha = \pi$).

Finally, we note that from
the general scaling relation for the edge exponent
\begin{equation} \label{sr}
\beta_e(\alpha) \, = \, \frac{\nu [d - 2 + \eta_{e \|}(\alpha)]}{2}
\, \, , \quad 2 \leq d \leq 4 \, \, ,
\end{equation}
which describes the singular temperature dependence of the 
order parameter at the edge,
one has within mean-field theory
[see Eq.\,(\ref{eta}), here $\eta = 0$ and $\nu = 1/2$]
\begin{equation} \label{beta1}
\beta_e(\alpha) \, = \, \frac{1 + \eta_{e \perp}(\alpha)}{2} 
\, \, , \quad d = 4 \, \, .
\end{equation}
In the limit $\alpha \to \pi$ corresponding to a planar surface
Eq.\,(\ref{beta1}) yields $\beta_e(\pi) = \beta_1 = 2$, 
in accordance with the
general scaling relation $\beta_1 = d \nu$ for $2 \leq d \leq 4$
at the normal transition for the semi-infinite geometry 
\cite{Die94}.


\subsection{Exact results in $d = 2$}
\label{secIVB}

In $d = 2$ rigorous results for the edge exponent 
$\eta_{e \perp}(\alpha)$ can be obtained  starting 
from the correlation function in 
the half-plane at the normal transition \cite{Car84}
\begin{equation} \label{G2d}
G(x,y;x',y') = (y y')^{-\eta/2} \ {\cal G}
\left( {(x - x')^2 + y^2 + y'^2 \over y y'} \right)\; ,
\end{equation}
where $x,x'$ are the coordinates of the two points
along the surface and $y,y'$ are their distances
from the surface \cite{Car84}.
Applying the conformal mapping and carrying out the limit $r'/r \to \infty$
at fixed $r > 0$ leads to
\begin{equation} \label{G2dwedge}
G(r,\theta;r',\theta') \sim \Big( {\pi \over \alpha} \Big)^{\eta}
\Big[ \sin\left({\pi \over \alpha} \theta\right)
\sin\left({\pi \over \alpha} \theta'\right)
\Big]^{(\eta_\| - \eta) / 2} \ (r r')^{-\eta/2} \
\Big( {r' \over r} \Big)^{-\pi \eta_\| / (2\alpha)}
\end{equation}
for the correlation function in the wedge.
From Eq.\,(\ref{G2dwedge}) one can read off the
scaling relation \cite{Car84}
\begin{equation} \label{eta2d}
\eta_{e \perp}(\alpha) \, = \, 
{\eta \over 2} \, + \, {\pi \over \alpha}\, {\eta_\| \over 2} \, \, .
\end{equation}
In the limit $\alpha \to \pi$ 
corresponding to a planar surface one has  
$\eta_{e \perp}(\pi) = (\eta + \eta_\|)/2 = \eta_{\perp}$ 
as expected [see Eq.\,(\ref{eta})].
In $d = 2$ the edge consists of
a single point so that one cannot define correlations along
the edge. However, if we formally define $\eta_{e \|}(\alpha)$ by
Eq.\,(\ref{eta}) we find 
$\eta_{e \|}(\alpha) = (\pi/\alpha) \eta_\|$. 
From $\beta_1 = 2$ \cite{Die94} we obtain $\eta_\| = 4$ at the
normal transition within the Ising universality class,
which implies [see Eq.\,(\ref{sr}), here $\nu = 1$]
\begin{equation} \label{beta12d}
\beta_e(\alpha) \, = \, 
\frac{\eta_{e \|}(\alpha)}{2} \, = \frac{2 \pi}{\alpha}
\, \, , \quad d = 2 \, \, ,
\end{equation}
for the edge exponent of the magnetization. 
We note that $\beta_e$ given by 
Eq.\,(\ref{beta12d}) is four times larger than its corresponding
value at the ordinary transition \cite{Car83,GT84,BPP84}.


\section{Summary}
\label{summary}

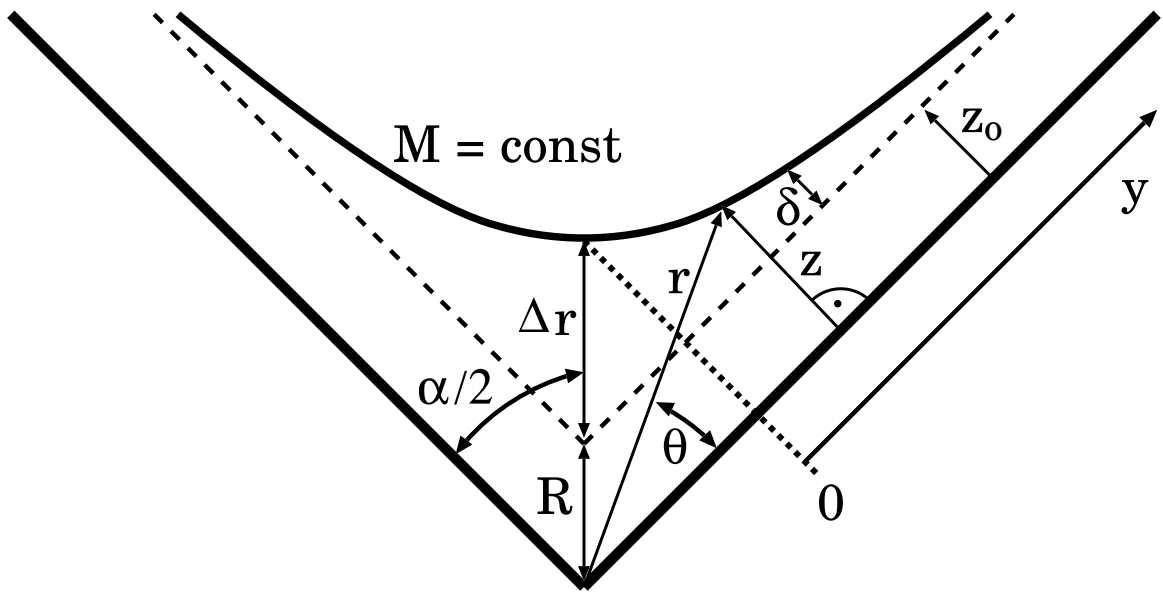
\begin{figure}
\caption{Cross section of a wedge perpendicular to its edge with 
opening angle $\alpha$. The system is translationally invariant 
in the $(d-2)$-dimensional subspace parallel to the edge. The curve
labeled ``M = const'' represents a contour line of the order parameter
profile (see Sec.\,\ref{secIIIB}).
$r$ and $0 \le \theta \le \alpha$ are cylindrical coordinates.
$z = r \sin \theta$ is the normal distance from the surface of 
the wedge. The dashed lines are the asymptotes of the contour
line M = const which are a distance $z_0$ apart from the surface.
$R = z_0 / \sin(\alpha/2)$ is the distance between the intersection
of the asymptotes and the edge of the wedge. 
$\Delta r + R$ is the distance between the contour line in the center
and the edge. The contour line approaches the asymptotes such that
$\delta(y \to \infty) \to 0$ and 
$\delta(y = 0) = \Delta r \sin(\alpha/2)$ where $y$ measures 
the distance along the wall.}
\label{fig_wedge}
\end{figure}

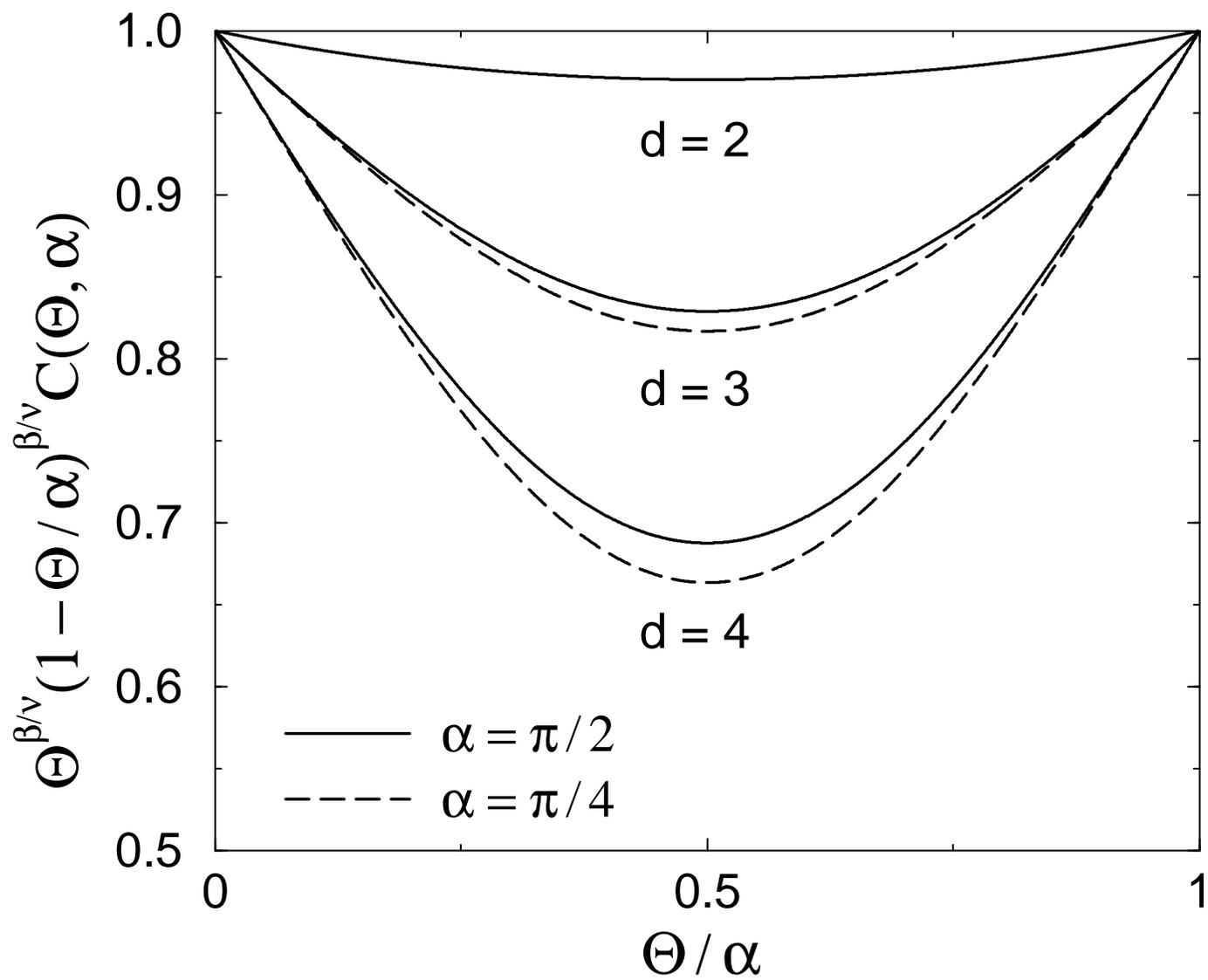
\begin{figure}
\caption{Universal scaling function ${\cal C}(\theta; \alpha)$ 
[see Eq.\,(\ref{mpmmhalf})] as function of the polar 
angle $\theta$ for fixed opening angles $\alpha = \pi / 2$ 
and $\alpha = \pi / 4$ (compare Fig.\,\ref{fig_wedge}).
The divergences at $\theta = 0$ and $\theta = \alpha$ 
are split off [compare Eq.\,(\ref{theta})].
The curves for $d = 3$ show the linear interpolation
between the corresponding exactly known curves for $d = 2$ 
[see Eq.\,(\ref{function})] and $d = 4$ [see Eq.\,(\ref{pi})].
Note that the shape of the curve for $d=2$ is independent 
of $\alpha$ for $0 < \alpha < \pi$.}
\label{fig_C}
\end{figure}

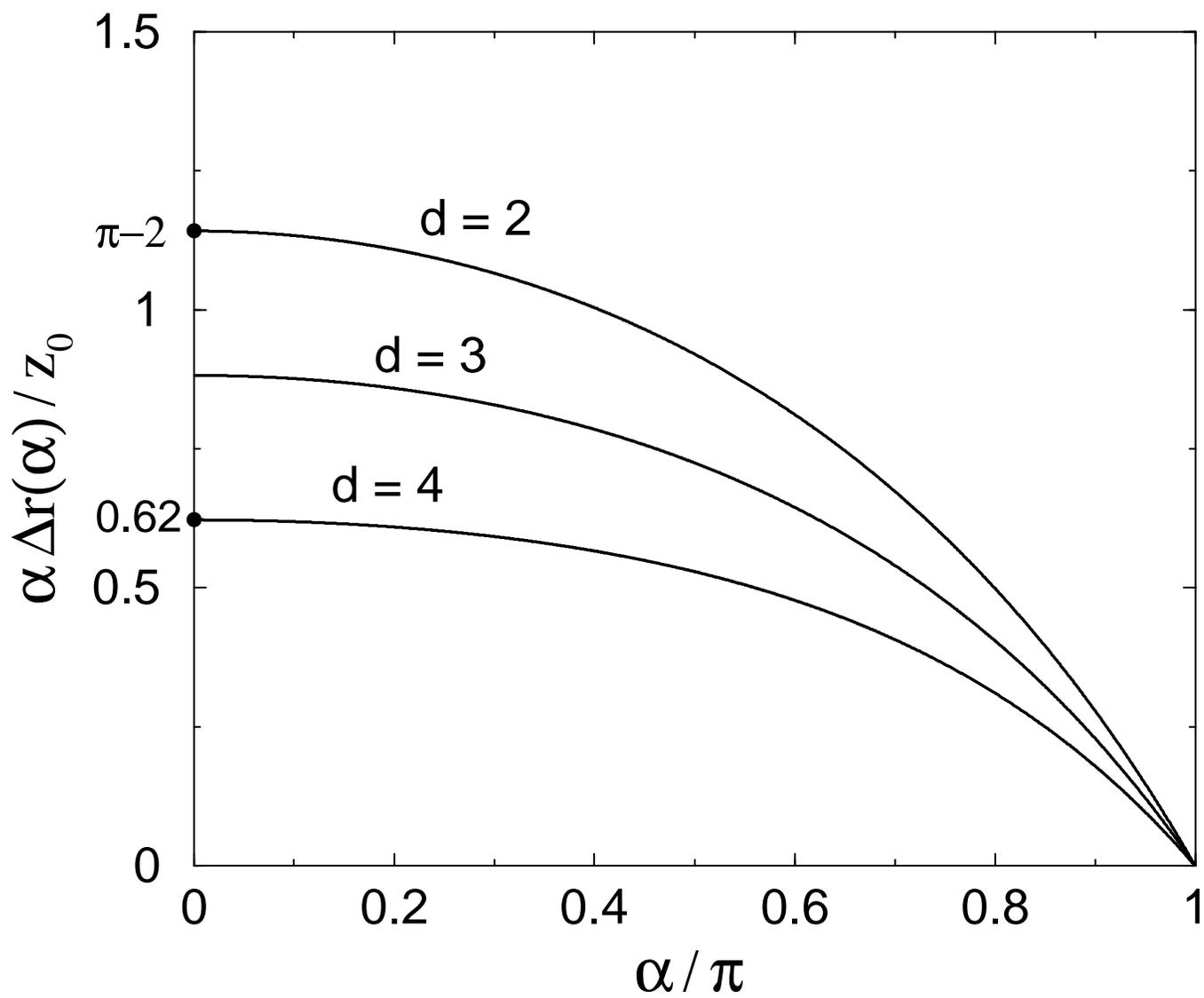
\begin{figure}
\caption{Universal dependence of the deviation $\Delta r = r - R$
in units of the reference distance $z_0$ 
of the contour line at the center of the wedge 
(compare Fig.\,\ref{fig_wedge}) on the opening angle $\alpha$.
The curve for $d = 3$ shows the linear interpolation
between the exactly known curves for $d = 2$ 
[see Eq.\,(\ref{Dr2d})] and $d = 4$ [see Eq.\,(\ref{Dr})].
$\Delta r$ is multiplied by $\alpha$ so that the product 
$\alpha \Delta r$ attains a finite value for $\alpha \to 0$
[see Eq.\,(\ref{Dr2d0}) for $d = 2$ and 
Eq.\,(\ref{Dr0}) for $d = 4$; 
we assume in addition that the power law 
$\Delta r(\alpha \to 0) \sim \alpha^{-1}$ remains valid 
for $2 < d < 4$].}
\label{fig_dr}
\end{figure}

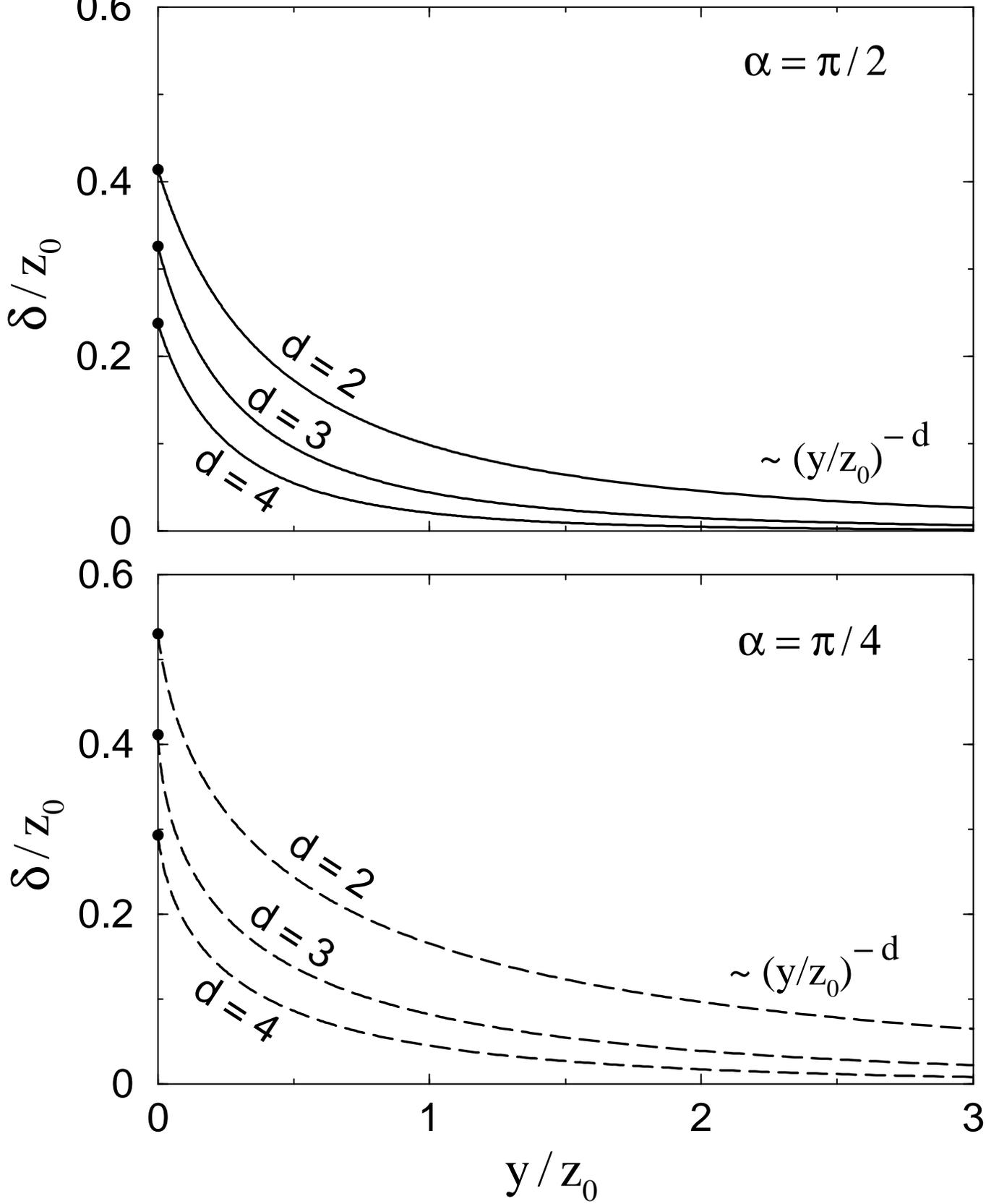
\begin{figure}
\caption{Universal dependence of the displacement $\delta(y;\alpha)$ 
of the contour line from its asymptote on the distance $y$ 
[compare Fig.\,\ref{fig_wedge} 
and see Eqs.\,(\ref{eins}) and (\ref{y})]
for fixed opening angles $\alpha = \pi / 2$ and $\alpha = \pi / 4$.
The curves for $d = 3$
correspond to a suitable interpolation [see Eq.\,(\ref{d3})]
between the exactly known curves for $d = 2$ and $d = 4$ 
so that for $y/z_0 \to \infty$ they exhibit the correct asymptotic 
decay $\sim (y/z_0)^{-3}$ in accordance with the 
distant-wall correction.}
\label{fig_delta}
\end{figure}

We have investigated the universal local properties of the order 
parameter profile in a wedge with opening angle $\alpha$ 
(see Fig.\,\ref{fig_wedge}) for the normal transition.  
We have obtained the following main results:
\begin{itemize}
\item[1.]Near $T_c$ the order parameter is determined by universal
scaling functions and the two nonuniversal bulk amplitudes
$a$ and $\xi_0^{+}$
[see Eq.\,(\ref{mpm})]. At $T_c$ the order parameter profile reduces to a
power law $r^{-\beta/\nu}$ in radial direction multiplied by a
universal amplitude function depending on the polar angle $\theta$ and
the opening angle $\alpha$
[see Eqs.\,(\ref{mpmTc}) and (\ref{mpmmhalf})]. 
The amplitude function is symmetric around the midplane and diverges 
as $\theta^{-\beta/\nu}$ upon approaching the surfaces forming the 
wedge [see Eq.\,(\ref{theta})].
\item[2.]We have determined the universal amplitude function 
${\cal C}(\theta, \alpha)$ within mean-field theory, i.e., for space
dimension $d=4$ [see Eqs.\,(\ref{pi}) and (\ref{2pi})], where
the order parameter profile in the wedge and for $T = T_c$ 
can be obtained from the order parameter profile in the 
film geometry for $T < T_c$ [see Eq.\,(\ref{C2})].
In conjunction with
exact results in $d=2$ [see Eq.\,(\ref{function})] we construct
an estimate for ${\cal C}(\theta, \alpha)$ for $d=3$ 
(see Fig.\,\ref{fig_C}).
\item[3.]The amplitude function determines the meniscus-like contour
lines of a constant value of the critical order parameter profile
[see Eq.\,(\ref{contour}) and Fig.\,\ref{fig_wedge}]. The deviation
$\Delta r$ 
of the contour line relative to its asymptotes 
(compare Fig.\,\ref{fig_wedge})
from the corner of the wedge vanishes linearly in the planar limit
$\alpha\to\pi$ [see Eqs.\,(\ref{Drpi}) and (\ref{Dr2dpi})] and diverges
$\sim\alpha^{-1}$ for $\alpha\to 0$ [see Eqs.\,(\ref{Dr0}) and
(\ref{Dr2d0})]. Figure \ref{fig_dr} presents an estimate 
of the function $\Delta r(\alpha)$ for $d=3$.
\item[4.]The contour lines approach their asymptotes
as $y^{-d}$ for increasing lateral distance $y$
[see Fig.\,\ref{fig_wedge} and Eq.\,(\ref{eins})].
This follows from an analysis of distant-wall
corrections in terms of the stress tensor (see Sec.\,\ref{secIIIC}).
The explicit results for $d=2$ and $d=4$ allow one to construct an
estimate for the corresponding behavior in $d=3$ 
(see Fig.\,\ref{fig_delta}).
\item[5.]The decay of the two-point correlation function at $T_c$ away
from the edge and parallel to the wedge is governed by the critical
edge exponents $\eta_{e\perp}(\alpha)$ and $\eta_{e\|}(\alpha)$,
respectively [see Eqs.\,(\ref{perp}) and (\ref{para})]. 
They fulfill the scaling relation 
$\eta_{e\perp}(\alpha)=[\eta+\eta_{e\|}(\alpha)]/2$
with the bulk exponent $\eta$. Based on the
quite accurate estimate in 
$d=4$ [see Eq.\,(\ref{etaperp}) and Appendix B] and the 
exact result in $d=2$
[see Eq.\,(\ref{eta2d})] Fig.\,\ref{fig_eta} presents an estimate for
$\eta_{e\perp}(\alpha)$ in $d=3$. Equations (\ref{beta1}) and
(\ref{beta12d}) provide the critical edge exponent $\beta_e(\alpha)$ of
the order parameter for the normal transition.
Equation (\ref{Sspectral}) gives the full structure 
factor for the wedge geometry and for $T = T_c$ within 
mean-field approximation.

\end{itemize}


\section*{Acknowledgements}
We thank G. Sommer for a helpful collaboration at an early 
stage of this work. M.K. gratefully acknowledges 
support by the German Science Foundation through a Heisenberg 
Stipendium. The work of A.H. and S.D. has been supported by the 
German Science Foundation through Sonderforschungsbereich 237 
{\em Unordnung und gro{\ss}e Fluktuationen\/}.


\appendix

\section{Elliptic functions}
\label{appA}

Here we summarize a few properties of elliptic functions as far
as they are needed for the derivation of the results obtained in 
this paper. For further information we refer to the literature
(see, e.g., Refs.\,\cite{Law89,GR65,Kam71}). The properties of 
the Jacobian elliptic functions can be derived starting from
the Jacobi amplitude $\mbox{am}(u;k)$ which is implicitly 
defined by the incomplete elliptic integral of the first kind:
\begin{equation} \label{EllipInt1}
u \, = \, \int_0^{\textstyle{\mbox{am}(u;k)}}
{d \varphi \over \sqrt{1 - k^2 \sin^2 \varphi}} \, \, .
\end{equation}
The complete elliptic integral of the first kind $K = K(k)$ 
is defined by $\mbox{am}(K;k) = \pi/2$. For the 
derivation of Eq.\,(\ref{Drpi}) we quote the expansion of $K(k)$ 
in powers of the modulus $k$:
\begin{equation} \label{Kkseries}
K(k) = {\pi \over 2} \left[ 1 + {1 \over 4} k^2 + {9 \over 64} k^4
+ {\cal O}(k^6) \right] \, \, .
\end{equation}
From the first derivative of Eq.\,(\ref{EllipInt1}) with respect 
to $u$ one obtains for $k^2 \leq 1$ the relation
\begin{equation} \label{dn}
\mbox{dn}(u;k) \equiv \frac{\partial}{\partial u} 
\mbox{am}(u;k) = \sqrt{1 - k^2 \mbox{sn}^2(u;k)}
\end{equation}
for the delta amplitude $\mbox{dn}(u;k)$ using 
the standard notations
\begin{equation} \label{sncn}
\mbox{sn}(u;k) \equiv \sin[\mbox{am}(u;k)] \, \, ,
\quad \mbox{cn}(u;k) \equiv \cos[\mbox{am}(u;k)] \, \, .
\end{equation}
Due to $\mbox{am}(0;k) = 0$ and $\mbox{am}(K;k) = \pi/2$ one has
\begin{mathletters} 
\label{special}
\begin{equation} \label{speciala}
\mbox{sn}(0;k) = 0 \, \, , \quad
\mbox{cn}(0;k) = 1 \, \, , \quad 
\mbox{dn}(0;k) = 1 \, \, ,
\end{equation}
\begin{equation} \label{specialb}
\mbox{sn}(K;k) = 1 \, \, , \quad
\mbox{cn}(K;k) = 0 \, \, , \quad
\mbox{dn}(K;k) = \sqrt{1 - k^2} \, \, .
\end{equation}
\end{mathletters}
\noindent
The derivation of Eq.\,(\ref{SDEmMF}) is based on the Taylor
expansions
\begin{mathletters} 
\label{sndnseries}
\begin{equation} \label{sndnseriesa}
\mbox{sn}(u;k) = u - {1 + k^2 \over 3!} u^3 + {1 + 14k^2 + k^4 \over 5!} u^5
- {1 + 135 k^2 + 135 k^4 + k^6 \over 7!} u^7 + {\cal O}(u^9) \, \, ,
\end{equation}
\begin{equation} \label{sndnseriesb}
\mbox{dn}(u;k) = 1 - {k^2 \over 2} u^2 + {k^2 (4 + k^2) \over 4!} u^4
- {k^2 (16 + 44 k^2 + k^4) \over 6!} u^6 + {\cal O}(u^8) \, \, .
\end{equation}
\end{mathletters}
\noindent
Finally, we quote the relation
\begin{equation} \label{EllipInt2}
E(\mbox{am}(u;k),k) = \int_0^u \mbox{dn}^2(x;k) dx =
\int_0^{\textstyle{\mbox{am}(u;k)}} \sqrt{1 - k^2 \sin^2 \varphi}\ d \varphi
\end{equation}
between the incomplete elliptic integral of the second kind $E(x,k)$ and the
delta amplitude $\mbox{dn}(u;k)$. $E(k) \equiv E(\pi/2,k)$ is the
complete elliptic integral of the second kind.

As demonstrated in Appendix A of Ref.\,\cite{Kre97} the mean-field order parameter
profile can be obtained from the observation that 
$[{\cal C}(\theta;\alpha)]^2$
is a Weierstrass $\wp$-function up to an additive constant. The Weierstrass
$\wp$-function is an elliptic function which is related to squares of certain
Jacobian elliptic functions \cite{Law89,GR65}. 
It solves the differential equation
\begin{equation} \label{WPeq}
[\wp'(u)]^2 = 4 \wp^3(u) - g_{2\,} \wp(u) - g_3
\end{equation}
where $g_2$ and $g_3$ are the invariants of $\wp$. Note that no term
quadratic in $\wp$ appears on the rhs of Eq.\,(\ref{WPeq}). 
This condition determines the
additive constant in the relation between $\wp$ and ${\cal C}^2$ which
appears as the ``potential'' in the eigenvalue problem in Eq.\,(\ref{psit}). 
For the derivation of the spectrum $\{ \lambda_n \}$ (see Appendix B) 
the Weierstrass $\zeta$-function is also needed. It is the negative 
integral of $\wp(u)$ and can be written as
\begin{equation} \label{Wzeta}
\zeta(u) = {1 \over u} - \int_0^u \left[ \wp(z) - 
{1 \over z^2} \right] dz \, \, .
\end{equation}
Note that $\zeta(u)$ is {\em not} an elliptic function. For the explicit
calculation of the spectrum $\{ \lambda_n \}$ we finally quote the 
Laurent series of $\wp(u)$ and $\zeta(u)$ around $u = 0$:
\begin{mathletters} 
\label{Wseries}
\begin{equation} \label{Wseriesa}
\zeta(u) = {1 \over u} - {g_2 \over 60} u^3 - {g_3 \over 140} u^5
\, + \, {\cal O}(u^7) \, \, ,
\end{equation}
\begin{equation} \label{Wseriesb}
\wp(u) = {1 \over u^2} + {g_2 \over 20} u^2 + {g_3 \over 28} u^4
\, + \, {\cal O}(u^6) \, \, .
\end{equation}
\end{mathletters}


\section{Eigenmode spectrum}
\label{appB}

The spectrum of the eigenvalue problem defined by Eq.\,(\ref{psit}) 
can be determined along similar lines as in Appendix B of Ref.\,\cite{Kre97}. 
The Weierstrass function associated with ${\cal C}(\theta;\alpha)$ 
[see Eq.\,(\ref{pi})] can be written in the form
$\wp(\theta) = {\cal C}^2(\theta;\alpha) - a$,
where $a$ is a constant to be determined. 
Based on Eq.\,(\ref{C2}) we aim at obtaining a
differential equation for $\wp$ which is 
of the form given by Eq.\,(\ref{WPeq}). This 
requirement is fulfilled if $a = 1/3$ which results in 
\begin{mathletters} 
\label{WPwedge}
\begin{equation} \label{WPwedgea}
\wp(\theta) = \left[ {2K \over \alpha} {\mbox{dn}(2K \theta / \alpha;k)
\over \mbox{sn}(2K \theta / \alpha;k)} \right]^2 \, - \, {1 \over 3} \, \, , 
\end{equation}
\begin{equation} \label{WPwedgeb}
g_2 = {4 k^2 (1 - k^2) \over (1 - 2 k^2)^2} \, + \, {4 \over 3} \, \, , 
\end{equation}
\begin{equation} \label{WPwedgec}
g_3 = {4 k^2 (1 - k^2) \over 3 (1 - 2 k^2)^2} \, + \, {8 \over 27} \, \, . 
\end{equation}
\end{mathletters}
\noindent
The eigenvalue problem in Eq.\,(\ref{psit}) can now be cast into
a Lam{\'e} equation \cite{Kam71}
\begin{equation} \label{Lame}
-\psi_n''(\theta) \ + \ 6 \ \wp(\theta) \psi_n(\theta) = 
\epsilon_n \ \psi_n(\theta) 
\end{equation}
where $\epsilon_n = \lambda_n - 2$. As discussed in 
Appendix B in Ref.\,\cite{Kre97} the eigenvalue spectrum 
is given by the solution of the two equations
\begin{mathletters} 
\label{EVeq}
\begin{equation} \label{EVeqa}
2 a_n \zeta(\alpha/2) - \alpha \left[\zeta(a_n) + {\wp'(a_n) \over 2\wp(a_n)
- \epsilon_n/3} \right] = n \pi i \, \, ,
\end{equation}
\begin{equation} \label{EVeqb}
\wp(a_n) = {\epsilon_n^3 - 27 g_3 \over
27 g_2 - 9 \epsilon_n^2} \, \, .
\end{equation}
\end{mathletters}
\noindent
The mode numbers $n$ are integers (see below) and $a_n$ are 
auxiliary parameters with the property
$a_n \to 0$ for $n \to \infty$. For large $n$
Eq.\,(\ref{EVeq}) can be solved asymptotically by using the expansions
quoted in Eq.\,(\ref{Wseries}). From the expansion for $\wp(a_n)$ we obtain
for large $n$
\begin{equation} \label{epsexp}
\epsilon_n = -{9 \over a_n^2} \left[ 1 + {7 g_2 \over 540} a_n^4
+ {\cal O}(a_n^6) \right]
\end{equation}
which implies the expansion
\begin{equation} \label{aexp}
{2 \over \alpha} \ \zeta(\alpha/2) a_n - {3 \over a_n} - {g_2 \over 180} a_n^3
 + {\cal O}(a_n^5) = {n \pi i \over \alpha} \, \, .
\end{equation}
Equation (\ref{aexp}) can be solved with the ansatz 
$a_n = 3\alpha i /(n\pi) [1 + A n^{-2} +B n^{-4} + {\cal O}(n^{-6})]$.
Inserting the solution into Eq.\,(\ref{epsexp}) leads to
\begin{equation} \label{epsn}
\epsilon_n = \left( {n \pi \over \alpha} \right)^2 \left\{ 1 -
{12 \over (n \pi)^2} \ \alpha \ \zeta(\alpha/2) - {36 \over (n \pi)^4} \left[
\left( \alpha\ \zeta(\alpha/2) \right)^2 - {g_2 \over 48} \alpha^4 \right]
+ {\cal O}(n^{-6}) \right\} \, \, ;
\end{equation}
$\lambda_n = \epsilon_n + 2$ yields the desired spectrum.

The allowed mode numbers $n$ can be obtained by considering the
special case $\alpha = \pi$ corresponding to a planar surface,
for which ${\cal C}(\theta;\pi) = 1/\sin \theta$ 
and Eq.\,(\ref{psit}) can be solved in closed form. 
For the present problem the
Weierstrass $\zeta$-function can directly be derived from
Eqs.\,(\ref{Wzeta}) and (\ref{WPwedge}) so that
\begin{equation} \label{Wzetawedge}
\zeta(\theta) = {2K \over \alpha} \left[ {\mbox{dn}(2K\theta/\alpha;k) \over
\mbox{sn}(2K\theta/\alpha;k)}\ \mbox{cn}(2K\theta/\alpha;k)
+ E(\mbox{am}(2K\theta/\alpha;k),k) + {k^2 - 2 \over 3}\ {2K \over \alpha}\
\theta \right]
\end{equation}
with $K = K(k)$ as defined in Appendix A.
Equation (\ref{Wzetawedge})
explicitly demonstrates that $\zeta(\theta)$ is not an elliptic function. 
At the midplane $\theta = \alpha/2$ Eq.\,(\ref{Wzetawedge}) 
reduces to [see Eq.\,(\ref{special})]
\begin{equation} \label{Wzetamid}
\zeta(\alpha/2) = {2 K(k) \over \alpha} 
\left[ E(k) + {k^2 - 2 \over 3} K(k) \right] \, \, .
\end{equation}
According to Eq.\,(\ref{k0pi}) the special case $\alpha = \pi$ corresponds
to $k = 0$ for which $\zeta(\pi/2) = \pi/6$ from Eq.\,(\ref{Wzetamid}) and
$g_2 = 4/3$ from Eq.\,(\ref{WPwedge}). From Eq.\,(\ref{epsn}) we infer
$\epsilon_n = n^2 - 2$, i.e., $\lambda_n = n^2$, which is indeed the correct
eigenvalue spectrum for ${\cal C}(\theta;\pi) = 1 / \sin \theta$.
The eigenfunctions are normalizable for $n \geq 3$.
Therefore $\lambda_3$ is the lowest eigenvalue for this problem. 

For a numerical solution of Eq.\,(\ref{EVeq}) using, e.g., the Newton method the
asymptotic spectrum given by Eq.\,(\ref{epsn}) for $n \geq 3$ provides excellent
initial values for the iteration. In fact, these initial values already are
within 0.1\% of the exact spectrum even for the ground state $n = 3$ if
$0.1 < \alpha/\pi \leq 1$. This implies that the mean-field expression for
$\eta_{e \perp}(\alpha)$ given by Eq.\,(\ref{etaperp}) is quite accurate if
$\alpha$ is not too small.



\end{document}